\documentclass[pra,twocolumn,showpacs,superscriptaddress]{revtex4}

\usepackage{amsmath}
\usepackage{amsfonts}
\usepackage{graphicx}

\font\bbfnt=msam10
\def\gsim{\,\hbox{\bbfnt\char'046}\,}
\def\lsim{\,\hbox{\bbfnt\char'056}\,}

\begin{document}

\title{Damped bounces of an isolated perfect quantum gas}
\author{S. Camalet}
\affiliation{Laboratoire de Physique Th\'eorique de la Mati\`ere Condens\'ee, UMR 7600, 
Universit\'e Pierre et Marie Curie, Jussieu, Paris-75005, France}
\date{Received: date / Revised version: date }
\begin{abstract}

The issue of the thermalization of an isolated quantum system is addressed by considering 
a perfect gas confined by gravity and initially trapped above a certain height. 
As we are interested in the behavior of truly isolated systems, we assume the gas is in a pure 
state of macroscopically well-defined energy. We show that, in general, for single-particle 
distributions, such a state is strictly equivalent to the microcanonical mixed state at the same energy. 
We derive an expression for the time-dependent gas density which depends on the initial gas 
state only via a few thermodynamic parameters. Though we consider non-interacting particles, 
the density relaxes into an asymptotic profile, but which is not the thermal equilibrium one 
determined by the gas energy and particle number. 

 \end{abstract} 

\pacs{03.65.-w, 05.30.-d, 05.70.Ln,  05.60.Gg}

\maketitle

\section{Introduction}

The dynamics of a quantum system under the influence of a heat bath has been widely 
studied in the last decades. Under these canonical conditions, the system relaxes into 
thermal equilibrium at the bath temperature, provided the coupling to the bath is weak 
enough \cite{QDS}. Relaxation under microcanonical conditions, assumed in 
thermodynamics, is far less understood. A possible definition of microcanical conditions 
might be an energy-conserving coupling to environmental degrees of freedom 
\cite{Mahler1,Mahler2}. As is well known, the measurement process is convincingly 
accounted for by such a coupling of the measurement apparatus to its environment: 
the coherences between the measured system eigenstates are generically destroyed 
whereas the populations remain constant \cite{Zurek, Endo1, Endo2}. It therefore seems 
difficult to regard the evolution under such conditions as a relaxation process. However, 
one can wonder whether physically relevant system degrees of freedom relax towards 
their thermodynamic equilibrium values. 
This question is also of interest for truly isolated systems, i.e., 
in absence of any coupling to environmental degrees of freedom.

This issue is especially relevant for the n-particle reduced density matrices of a many-body 
system. Recently, different boson and fermion systems have been studied numerically 
\cite{RYDO, KLA,MWNM,Nature}. In these works, the system is initially prepared in 
the ground state of a Hamiltonian different from that governing the time evolution and 
single-particle distributions are evaluated. The distributions considered were found to relax 
into thermal or non-thermal distributions irrespective of the integrable or non-integrable 
nature of the system. In \cite{PRL}, the Joule expansion of an isolated perfect quantum gas 
has been obtained analytically. In this study, the gas is initially trapped in a subregion 
of the entire accessible volume but is not assumed to be in the corresponding ground state. 
Following \cite{Tasaki,Mahler1,Mahler2,CT,EPJB}, a pure state of macroscopically 
well-defined energy has been considered and the thermodynamic limit expression for 
the time-dependent particle number density has been derived. Interestingly, for 
this single-particle distribution, the initial gas state is fully characterized by a few 
thermodynamic parameters, such as the gas energy and particle number 
or, equivalently, the corresponding microcanonical temperature and chemical 
potential defined from derivatives of the gas microcanonical entropy.    

In this paper, we consider 
an isolated quantum gas confined by a uniform gravitational field and a perfectly reflecting wall. 
The gas is at first trapped above a certain height and then left free to fall. Such experiments 
can be realized with laser-cooled atoms and bounces of both Bose-Einstein condensates 
and thermal clouds have been observed \cite{ENS,Bongs}. We study the case of a perfect gas. 
The model we examine is presented in the next section. 
In Sec.~\ref{Me}, we discuss the states of macroscopically well-defined energy. 
We show that, for single-particle distributions, such a pure state is strictly equivalent to 
the microcanonical mixed state at the same energy.  Our derivation is applicable to 
any many-body system. For a perfect gas confined by gravity, we are able to derive 
the time-dependent density profile in the limit of a large particle number. This is done 
in Sec.~\ref{Db}. We obtain damped oscillations of the gas density. However, 
the asymptotic particle number density of the gas is far from the thermal equilibrium one 
determined by the gas energy and particle number. We then compare the relaxation behavior 
of a truly isolated system to that of a system under the influence of an environment.
Finally, in the last section, we summarize our results and draw conclusions.

\section{Model \label{M}}  

We consider a one-dimensional perfect gas confined by a homogeneous gravitational field and 
an infinite potential wall. The $N$ particles of mass $m$ constituting the gas are described by 
the Hamiltonian
\begin{equation}
H_0 = \int_0^\infty dz \psi^{\dag} (z) \left[ -\frac{1}{2m} \partial_z^2 \psi (z) 
+ m g z \psi (z) \right] \label{H}
\end{equation}
where $\psi^\dag (z)$ creates a particle at position $z$ and $g$ is the acceleration of 
the gravitational field. Throughout this paper, we use units in which $\hbar=k_B=1$. 
The non-interacting Hamiltonian \eqref{H} can be easily generalized to higher dimensions. 
In the following, we treat in detail only the one-dimensional case but we also mention results 
for the 3D case. The Hamiltonian \eqref{H} can be characterised by the microscopic 
energy $\epsilon_0=(mg^2/2)^{1/3}$ and length $z_0=(2m^2 g)^{-1/3}$. For Rb atoms in 
the  Earth's gravitational field, $\epsilon_0$ and $z_0$ are, respectively, of the order of 
$10$ nK and $0.1 \; \mu$m. 
The single-particle eigenenergies and normalised eigenfunctions of $H_0$ are 
\begin{equation}
\epsilon_q= \epsilon_0 |a_q | \quad , \quad  \phi_q (z) 
= \frac{A(z/z_0+a_q)}{\sqrt{z_0} A'(a_q)} \Theta(z) 
\label{eigen}
\end{equation}
where $\Theta$ is the Heaviside step function, $A(x)=\int_0^\infty dt \cos(t^3/3+xt)/\pi$ 
the Airy function \cite{Af}, $A'$ its derivative and $a_q$ its zeros. We label these zeros by 
positive integers in the order of increasing absolute values. 
Initially, the gas is trapped above a certain height $Z$ and hence described 
by the Hamiltonian $H_Z$ given by the expression \eqref{H} with the lower limit $0$ 
replaced by $Z$. The single-particle eigenenergies and eigenfunctions of $H_Z$ are 
simply $\epsilon_{Z,k}=mgZ+\epsilon_k$ and $\phi_{Z,k}=\phi_k (z-Z)$. 

\section{Microcanonical evolution \label{Me}} 

As mentioned in the introduction, the gas is assumed to be initially in a pure state of 
macroscopically well-defined energy $E$ 
\begin{equation}
|\Psi \rangle = \sum_{|\alpha\rangle \in {\cal H}_E} \Psi_\alpha |\alpha \rangle  \label{psi}
\end{equation} 
where $|\alpha\rangle$ refers to the $N$-particle eigenstates of the Hamiltonian $H_Z$. 
The Hilbert space ${\cal H}_E$ is spanned by the states $|\alpha \rangle$ corresponding 
to eigenenergies in the interval $[E,E+\delta E]$ where $\delta E$ is much smaller than 
$E$ but much larger than the maximum level spacing of $H_Z$ in this interval. 
The subsequent time evolution of the gas is governed by the Hamiltonian $H_0$ 
which is different from $H_Z$. 
In this section, we first show that, in the large $N$ limit, the time-dependent gas density 
is the same for almost all states \eqref{psi} and equal to a microcanonical average 
at energy $E$ of the particle number density operator. We stress that this result 
does not rely on the specific form of the Hamiltonians $H_0$ and $H_Z$. 
Our derivation does not even have recourse to their non-interacting nature. 
We then discuss the particular Hamiltonian \eqref{H}. 
 
\subsection{Microcanonical typicality \label{Mt}}

The gas density is given by 
\begin{equation}
\rho(z,t) = \langle \Psi | e^{iH_0 t} \psi^{\dag} (z) \psi (z) e^{-iH_0 t} | \Psi \rangle . \label{rho}
\end{equation} 
To prove that $\rho$ is the same for almost all states \eqref{psi}, we use, following 
Refs. \cite{CT,EPJB,PRL}, the uniform measure on the unit sphere in ${\cal H}_E$
\begin{equation}
\mu \big( \{\Psi_\alpha\} \big) = \frac{(D-1)!}{\pi^D} \delta 
\Big( 1-\sum_{|\alpha\rangle \in {\cal H}_E} |\Psi_\alpha|^2 \Big) \label{mu}
\end{equation}
where $D$ is the dimension of ${\cal H}_E$. We assume that, for large $N$, 
this dimension is practically proportionnal to $\delta E$ and that 
a density of states can thus be defined as $n (E,N)=D/\delta E$ \cite{Diu}. 
This density satisfies Boltzmann's relation
\begin{equation}
\ln D = \ln \left[ n (E,N) \delta E \right] = S ( E , N )  \label{n}
\end{equation}  
where $S$ is the gas entropy. 

We now evaluate the Hilbert space average and variance of the number
\begin{equation}   
N_\delta (z,t) =  \int_z^{z+\delta} dz' \rho (z',t)
\end{equation}
following from the normalised distribution \eqref{mu}. The above length $\delta > 0$ can 
be as small as we want. For $N \gg 1$, the reduced distribution of a finite number $j$ of 
components $\Psi_{\alpha(1)}, \ldots, \Psi_{\alpha(j)}$, equals 
$\prod_{i=1}^{j} D\exp(-D | \Psi_{\alpha(i)} |^2)/\pi$. With this Gaussian distribution, 
we obtain the Hilbert space variance
\begin{equation}   
\overline{{N_\delta}^2} - \overline{N_\delta}^2  = \frac{1}{D^2} 
\sum_{|\alpha\rangle,  |\beta\rangle \in {\cal H}_E} 
\Big| \langle \alpha | \int_z^{z+\delta} dz'  \hat{\rho} (z',t) |\beta \rangle \Big|^2
\end{equation}
where $\hat{\rho} = \exp(iH_0t) \psi^{\dag} \psi \exp(-iH_0t)$ and 
$\overline{\phantom{|}\ldots\phantom{|}}$ denotes the average with respect to 
the measure \eqref{mu}. Upper bounding the sum over the states 
$|\beta\rangle \in {\cal H}_E$ by a sum over all the $N$-particle eigenstates $|\beta\rangle$ 
gives 
\begin{equation}
\overline{{N_\delta}^2} - \overline{N_\delta}^2 < \frac{1}{D} 
\int_z^{z+\delta} dz' \int_z^{z+\delta} dz''  \langle \hat{\rho} (z',t) \hat{\rho} (z'',t) \rangle \label{var}
\end{equation} 
where $\langle \ldots \rangle=\sum_{|\alpha \rangle \in {\cal H}_E} 
\langle \alpha | \ldots | \alpha \rangle/D$ is a microcanonical average at energy $E$. 
We reiterate that $| \alpha \rangle$ are the eigenstates of the Hamiltonian $H_Z$. 
For the Hilbert space average of the density $\rho$,  we find $\langle \hat{\rho} \rangle$. 
As the density-density correlation function on the right side of \eqref{var} is positive, 
the Hilbert space variance of $N_\delta$ is lower than $N^2/D$. 
Consequently, as $D = \exp(S) \sim \exp(N)$, this variance 
vanishes in the limit $N \gg 1$ and hence $\rho(z,t)=\langle \hat{\rho} (z,t) \rangle$ 
for almost all states \eqref{psi}. 

The above derivation is applicable to any single-particle distribution 
$\langle \Psi | e^{iH_0 t} c^{\dag}_p c^{\phantom{{\dag}}}_p e^{-iH_0 t} | \Psi \rangle$ 
where $c^{\dag}_{p} = \int dz \chi_p (z) \psi^{\dag} (z)$ and the wavefunctions $\chi_p$ 
form a basis of the single-particle Hilbert space. Moreover, since the only property 
of the observable $A=\int_z^{z+\delta} dz' {\hat \rho}$ used in this calculation 
is $\langle A^2 \rangle \ll \exp(N)$, the equivalence between the average 
$\langle \Psi | \ldots | \Psi \rangle$ and the microcanonical average $\langle \ldots \rangle$ 
shoud be valid for a large class of few-body observables. We remark that this equivalence 
is obtained here for a typical state $| \Psi \rangle$ not for an eigenstate $| \alpha \rangle$. 
Contrary to Refs. \cite{Tasaki, Srednicki}, we have not used any remarkable property of 
these eigenstates in our derivation.

\subsection{Perfect gas confined by gravity \label{Pgcg}}

For the non-interacting Hamiltonians $H_0$ and $H_Z$, it is useful to 
define the creation operators $c^\dag_{p} = \int dz \phi_p (z) \psi^\dag (z)$ and $c^\dag_{Z,p}$ 
given by similar expressions with $\phi_{Z,p}$ in place of $\phi_{p}$. 
With these definitions, the particle number density reads
\begin{equation}   
\rho (z,t)=\sum_{p,q} \langle c^{\dag}_{p} c^{\phantom{\dag}}_{q} \rangle \phi_p (z) \phi_q (z) 
e^{it(\epsilon_p-\epsilon_q)} \label{dens}
\end{equation}
where $\langle \ldots \rangle=\sum_{|\alpha \rangle \in {\cal H}_E} 
\langle \alpha | \ldots | \alpha \rangle/D$ as shown above. Here, an eigenstate 
$|\alpha \rangle$ corresponds to a set of occupation numbers $\{ n_k \}$ obeying 
$\sum_{k} n_k=N$. For fermions, $n_k$ is restricted to the values $0$ and $1$. The Hilbert 
space ${\cal H}_E$ is spanned by the states $|\alpha \rangle$ satisfying 
$E< \sum_{k} n_k \epsilon_{Z,k} <E+\delta E$. 
The expression \eqref{dens} can be further simplified using the following standard arguments 
\cite{Diu}. The microcanonical probability distribution of the occupation number $n_k$ is 
$P(n_k)={\hat n}(E-n_k \epsilon_{Z,k} , N - n_k )/n(E,N)$ where ${\hat n}$ is the density of 
states of the Hamiltonian $H_Z-\epsilon_{Z,k} c^{\dag}_{Z,k} c^{\phantom{\dag}}_{Z,k}$. 
This density obeys Boltzmann's relation \eqref{n} with the corresponding entropy ${\hat S}$. 
Expanding this entropy in the energy $n_k \epsilon_{Z,k} \ll E$ and number $n_k \ll N$ and 
taking into account that ${\hat S}=S$ in the large $N$ limit, result in 
$P(n_k) \propto \exp[-n_k(\epsilon_{Z,k}-\mu) /T]$ where the temperature $T$ and 
chemical potential $\mu$ are determined by 
\begin{equation}
\frac{1}{T} = \partial_E S (E,N) \quad , \quad  \frac{\mu}{T} = - \partial_N S (E,N) .  \label{Tmu} 
\end{equation}
The microcanonical averages in \eqref{dens} can thus be rewritten as 
\begin{equation}   
\langle c^{\dag}_{p} c^{\phantom{\dag}}_{q} \rangle = \sum_{k>0} 
\frac{\langle p | k \rangle\langle q | k \rangle}{e^{\epsilon_k/T}/\zeta \mp 1} \label{cpcq}
\end{equation}
where the upper sign is for bosons and the lower sign is for fermions, 
$\zeta=\exp[(\mu-mgZ)/T]$  and 
\begin{equation}   
\langle q | k \rangle = \int dz \phi_q(z) \phi_k (z-Z) 
= \frac{A(Z/z_0+a_q)}{A'(a_q) [a_q-a_k+Z/z_0]} . \label{spqk}
\end{equation} 
The second equality is derived in the Appendix.

Other choices than \eqref{psi} are possible to describe a gas state of macroscopic 
energy $E$. For example, due to the exponential $N$-dependence \eqref{n} of the density 
$n(E,N)$, one obtains the same gas density $\rho$ for typical states of ${\cal H}_E$ and of 
the Hilbert space spanned by the eigenstates $|\{ n_k \}\rangle$ satisfying $\sum_k n_k=N$ 
and $\sum_k n_k \epsilon_{Z,k} < E$. 
We also remark that, in three dimensions, 
for a free horizontal propagation, the number of particles $\rho(z,t)dz$ 
between $z$ and $z+dz$ is given by \eqref{dens} and \eqref{cpcq} with the thermal 
occupation function replaced by a sum over horizontal modes determined by the initial 
confining potential. If, for example, the gas is initially trapped in a square of side length 
$\sqrt{S} \gg (mT)^{-1/2}$, the thermal occupation factor in \eqref{cpcq} is replaced by 
$\mp mS T \ln[1\mp \zeta \exp(-\epsilon_k/T)]/2\pi$. In this case, the behavior of $\rho$ is 
qualitatively similar to that discussed in the following.

\section{Damped bounces \label{Db}}  

In this section, we derive the large $N$ limit expression for the time-dependent 
gas density given by \eqref{dens} and \eqref{cpcq}. 
We find that it relaxes to an asymptotic density profile $\rho_{\infty}(z)$. However, 
this density $\rho_\infty$ is not the thermal equibrium one determined by $E$ and $N$. 
It depends not only on these thermodynamic parameters but also on the height $Z$.

\subsection{Temperature, chemical potential and density profile of the initial state \label{Tcpdpis}}

\begin{figure}
\centering \includegraphics[width=0.45\textwidth]{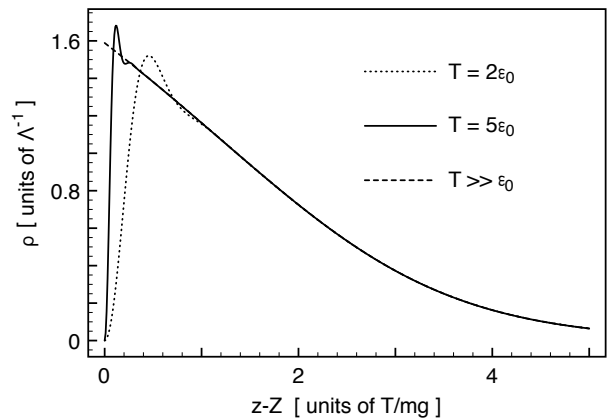}
\caption{\label{fig:rhoz0} Initial particle number density as a function of height for fermions, 
fugacity $\zeta=10$ and temperatures $T=2\epsilon_0$ and $5\epsilon_0$. The dashed 
line is the large $N$ limit. }
\end{figure}

The temperature $T$ and chemical potential $\mu$ are determined by the gas energy $E$ 
and particle number $N$ via \eqref{Tmu} or via the more convenient relations 
\begin{equation}   
N=\sum_{k>0} f_\zeta \left( |a_k|\frac{\epsilon_0}{T} \right) \quad , \quad 
E= \epsilon_0 \sum_{k>0} |a_k| f_\zeta \left( |a_k|\frac{\epsilon_0}{T} \right) \label{NE}
\end{equation}
where $f_\zeta(x)=[\exp(x)/\zeta \mp 1]^{-1}$. We remark that the fugacity $\zeta$ defined 
above and the temperature $T$ depend only on $N$ and $E$ and not on the height $Z$. 
As the large zeros of the Airy function are essentially given by $|a_k| \simeq (3\pi k/2)^{2/3}$, 
the density of levels $\epsilon_k=\epsilon_0 |a_k| $ increases with $k$ and the above sums 
are dominated by the contributions of zeros $|a_k| \sim T/\epsilon_0$ which are very close 
to their large-$k$ estimates. Moreover, for $T \gg \epsilon_0$, the sums \eqref{NE} can be 
approximated by integrals \cite{BEC} and hence 
\begin{equation}   
N=\left( \frac{T}{\epsilon_0} \right)^{3/2} I_{1/2} (\zeta) \quad , \quad 
E= \epsilon_0 \left( \frac{T}{\epsilon_0} \right)^{5/2} I_{3/2} (\zeta) \label{NEint}
\end{equation}
where $I_{\nu} (\zeta)=\int_0^\infty dx x^\nu f_\zeta(x)/\pi$. For a Bose gas, these expressions 
hold only for temperatures $T$ above the condensation temperature 
$T_B=\epsilon_0[N/I_{1/2}(1)]^{2/3}$.  Below this temperature, the right sides of the equalities 
\eqref{NEint} with $\zeta=1$ give the non-condensed fraction of the initial state and 
the corresponding energy. 

The initial particle number density $\rho(z,0)=\sum_{k>0} f_\zeta ( \epsilon_k/T ) \phi_k(z-Z)^2$ 
can be evaluated in a similar manner as follows. As the Airy function $A(x)$ vanishes rapidly 
with increasing $x$, the sum over $k$ is dominated by the terms $|a_k| \gsim (z-Z)/z_0$. 
For these terms, using the large negative $x$ approximation of $A(x)$ and 
$|A'(a_k)| \simeq \pi^{-1/2} |a_k|^{1/4}$, we find
\begin{equation}   
\phi_k(z-Z)^2 \simeq \frac{1+\sin[4[|a_k| - (z-Z)/z_0]^{3/2}/3]}{2 z_0 |a_k|^{1/2}[|a_k| - (z-Z)/z_0]^{1/2}} .
\end{equation}
The contribution of the sine term to $\rho(z,0)$ is negligible as it oscillates strongly with $k$ 
and we finally obtain the large $N$ expression 
\begin{equation}   
\rho(z,0) \simeq \frac{1}{\sqrt{\pi} \Lambda} \int_0^{\infty} \frac{dx}{\sqrt{x}}
f_\zeta \left[ x + \frac{mg}{T} (z-Z) \right] \Theta (z-Z) \label{rhoz0}
\end{equation}
where $\Lambda=(2\pi/mT)^{1/2} \propto (\epsilon_0/T)^{3/2} T/mg$ is the de Broglie 
thermal wavelength. The number of particles between $z>Z$ and $z+dz$ is equal to that of 
free particles in a box of volume $dz$ at temperature $T$ and chemical potential $\mu-mgz$. 
This expression does not describe the increase of $\rho$ in the small region $z-Z \lsim  \Lambda $ 
but is then very accurate, see Fig.\ref{fig:rhoz0}. The results shown in Fig.\ref{fig:rhoz0} are 
obtained by numerical summation over the exact eigenstates \eqref{eigen}. The characteristic 
length scale of the particle number density is thus essentially $T/mg$ for $\zeta < 1$. This length 
is of the order of $1$ mm for Rb atoms at $0.1$ mK in the Earth's gravitational field. As $\zeta$ 
and $T$ are independent of $Z$, the thermal equilibrium density profile of the gas described by 
the Hamiltonian $H_0$, energy $E$ and particle number $N$ is given by \eqref{rhoz0} with $Z=0$.

In the classical Maxwell-Boltzmann limit, $\Lambda \ll T/mgN$, the expression \eqref{rhoz0} 
simplifies to $\rho(z,0) \propto \exp[mg(Z-z)/T]$. This exponential decrease is always valid for 
large $z$. For a Fermi gas, in the opposite limit, the expressions \eqref{NEint} and \eqref{rhoz0} 
hold for a Fermi temperature $T_F = \epsilon_0 (3\pi N/2)^{2/3} \gg \epsilon_0$. In this case, 
\eqref{rhoz0} becomes $\rho(z,0) \propto [1+mg(Z-z)/T_F]^{1/2}$. For a Bose gas below 
the temperature $T_B$, the density profile of the non-condensed gas is given by \eqref{rhoz0} 
with $\zeta=1$ whereas the particle number density of the condensate is proportionnal to 
$A(z/z_0+a_1)^2$ which practically vanishes for $z \gsim z_0 = \epsilon_0/mg$. For 
$T \lsim T_B$, the ratio of the sizes of the condensate and thermal cloud is of the order of 
$N^{-2/3}$. We remark that though \eqref{rhoz0} with $\zeta=1$ diverges in the limit 
$mg(z-Z)/T \rightarrow 0$, the number of non-condensed bosons between $Z$ and  
$Z+z_0$ is negligible. 

\subsection{Density relaxation \label{Dr}}

\begin{figure}
\centering \includegraphics[width=0.45\textwidth]{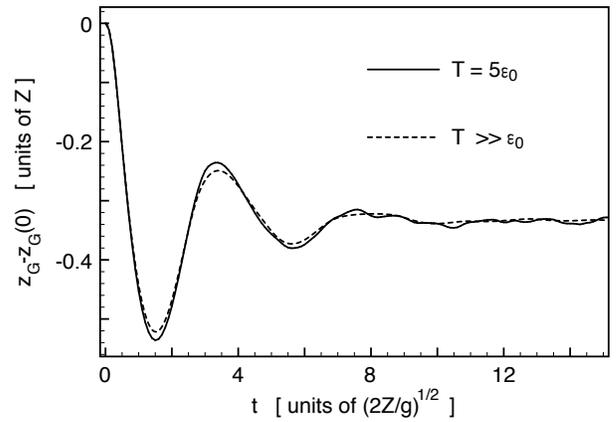}
\caption{\label{fig:zGtT} Gas center of mass as a function of time for fermions, fugacity 
$\zeta=10$, temperature $T=5\epsilon_0$ and $Z=T/mg$. The dashed line is 
the large $N$ limit. This approximation is excellent already for $T \simeq 10\epsilon_0$.}
\end{figure}

The time evolution of the density \eqref{dens} can be characterized by the motion of 
the center of mass 
\begin{equation}   
z_G (t) = N^{-1} \int_0^\infty dz z \rho (z,t) . \label{zGdef}
\end{equation}
As detailed in the Appendix, by evaluating the second order derivative of $z_G$ and 
using the property $\partial_z \phi_q (0^+) = (2m^2 g)^{1/2}$, we find 
\begin{eqnarray}   
z_G (t) &=& z_G (0) + g \int \frac{d\omega}{\omega^2} J(\omega) [1-\cos(\omega t)]  \nonumber \\
J(\omega) &=& N^{-1} \sum_{p \ne q} \langle c^{\dag}_{p} c^{\phantom{\dag}}_{q} \rangle 
\delta(\omega - \epsilon_p + \epsilon_q ) . \label{zG}
\end{eqnarray}
To determine the large $N$ limit of the function $J$, we first rewrite it as 
\begin{eqnarray}   
J(\omega) &=& N^{-1} \int_0^\infty dz \int_0^\infty dz' \Omega(z,z') \label{J} \\
&& \times \sum_{p \ne q} \delta(\omega - \epsilon_p + \epsilon_q ) \phi_p (z+Z) \phi_q (z'+Z) 
\nonumber \\
\Omega(z,z') &=& \sum_{k >0} f_\zeta ( \epsilon_k/T ) \phi_k(z) \phi_k(z') \\
&\simeq& \frac{1}{ \Lambda} \int dx
f_\zeta \left[ \pi x^2 + \frac{mg}{T} z \right] e^{i2 \pi x (z-z')/\Lambda} . \nonumber
\end{eqnarray} 
The approximate expression for $\Omega$ holds for $|z-z'| \lsim \Lambda$. For larger 
$|z-z'|$, $\Omega$ practically vanishes. The function $\Omega$ is simply related to 
the single-particle density matrix at initial time 
$\langle \psi^\dag (z) \psi (z') \rangle=\Omega(z-Z,z'-Z)$, see \eqref{rhoz0}. Due to 
the factor $\Omega(z,z')$, the main contribution to the integral \eqref{J} comes from 
the region $z \sim T/mg$ and $|z-z'| \lsim \Lambda$. The sum in the expression \eqref{zG} 
is thus dominated by the terms for which 
$\epsilon_p, \epsilon_q \sim T$ and $|a_p -a_q| \lsim ( \epsilon_0 / T)^{1/2}$. For these terms, 
$\epsilon_p - \epsilon_q \simeq \pi \epsilon_0 (p-q) ( \epsilon_0 / \epsilon_q)^{1/2}$. We use 
this approximation and rewrite the sum in \eqref{J} as a sum over $q$ and $s=p-q$. 
For $T \gg \epsilon_0$, the sum over $q$ is well approximated by an integral and the second 
factor in the integrand in \eqref{J} is found to vanish for $|z-z'| \gg \Lambda$. Consequently, 
the lower limit of the integral over $z'$ can be pushed to $-\infty$ and we obtain 
\begin{eqnarray}   
z_G (t) &\simeq&  z_G (0) + Z \frac{8 \pi {\bar Z}^{3/2}}{I_{1/2}(\zeta)} \sum_{s > 0} (-1)^s s^2 
\int_0^{s \pi} \frac{dx}{x^6}   \label{zGfinal} \\ 
&& \times  \frac{\sin  \sqrt{ (s\pi)^2 - x^2 }}{e^{{\bar Z} [ ( s\pi/x )^2 - 1]} / \zeta \mp 1} 
\left[ 1-\cos \left( x \sqrt{ \frac{g}{2 Z} } t \right) \right]  \nonumber
\end{eqnarray}
where ${\bar Z}=mgZ/T$. We remark that, similarly to Ref.\cite{PRL}, it remains a sum over 
a discrete index in this large $N$ expression. For $Z \sim T/mg$, the time $(Z/g)^{1/2}$ is 
of the ordrer of $10$ ms for Rb atoms at $0.1$ mK in the Earth's gravitational field. To better 
apprehend the characteristics of the many-body case studied here, it is instructive to compare 
with the case of a single particle. The average position of a particle can be written in the form 
\eqref{zG}-\eqref{J} with $Z=0$ and $\Omega(z,z')$ replaced by $\varphi (z)^* \varphi (z')$ 
where $\varphi$ is the initial wave function of the particle. This density matrix is very different 
from $\Omega$. Its extents along and perpendicular to the diagonal $z'=z$ are the same 
whereas $\Omega$ is almost diagonal which plays an essential role in the above derivation.

\begin{figure}
\centering \includegraphics[width=0.45\textwidth]{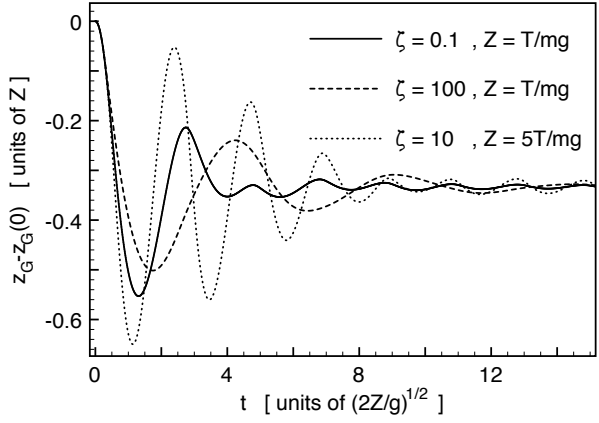}
\caption{\label{fig:zGtphiZ} Gas center of mass as a function of time in the large $N$ limit 
for fermions and $(\zeta,mgZ/T)=(0.1,1)$, $(100,1)$ and $(10,5)$.}
\end{figure}

\begin{figure}
\centering \includegraphics[width=0.45\textwidth]{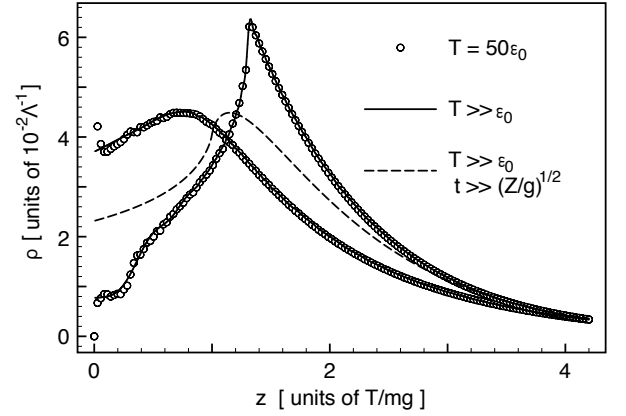}
\caption{\label{fig:rhozt} Gas density as a function of height for fermions, $\zeta=0.1$, 
$Z=T/mg$, $T=50\epsilon_0$ and times $t=1.3 (2Z/g)^{1/2}$ 
and $t=2.7 (2Z/g)^{1/2}$. These times correspond to the first minimum and the second 
maximum of the gas center of mass, see Fig. \ref{fig:zGtphiZ}. The full lines are 
the large $N$ density at the same times. The dashed line is the large $N$ density at infinite 
time.}
\end{figure}

\begin{figure}
\centering \includegraphics[width=0.45\textwidth]{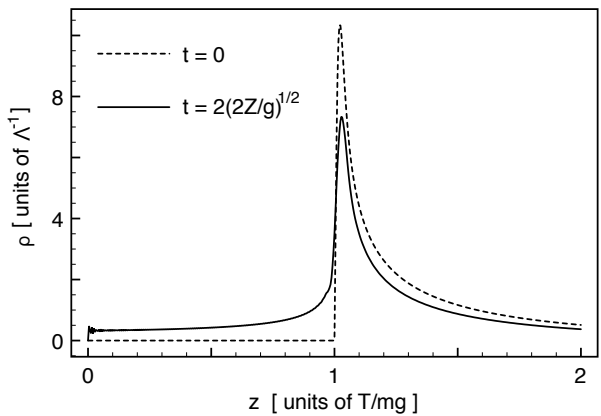}
\caption{\label{fig:BEC} Gas density as a function of height for bosons, condensate fraction 
of $0.1$, $Z=T/mg$, $T=50\epsilon_0$ and times $t=0$ and 
$t=2 (2Z/g)^{1/2}$.}
\end{figure}

In the limit $t \gg (Z/g)^{1/2}$, the contribution of the cosine term in \eqref{zGfinal} vanishes. 
This expression thus describes the relaxation of the center of mass $z_G$ to an asymptotic 
value $z_G(\infty)$. From \eqref{rhoz0} and \eqref{rhoinfty}, it can be shown that 
$z_G(\infty)=z_G(0)-Z/3$. A gravitational potential energy of only $NmgZ/3$ is converted into 
kinetic energy whereas the potential energy difference between the initial density \eqref{rhoz0} 
and the thermal equilibrium one given by \eqref{rhoz0} with $Z=0$ is $NmgZ$. The function 
\eqref{zGfinal} is non-monotonic, the motion of the gas center of mass consists of damped 
oscillations, see Figs. \ref{fig:zGtT} and \ref{fig:zGtphiZ}. The results shown in Figs \ref{fig:zGtT}, 
\ref{fig:rhozt} and \ref{fig:BEC} are obtained by numerical evaluation of the sums \eqref{zG} 
and \eqref{dens} with the second form of the scalar product \eqref{spqk}. For finite $T/\epsilon_0$, 
the motion of the gas center of mass is not strictly a relaxation to the asymptotic value 
$z_G(\infty)$. The distance $z_G(t)-z_G(\infty)$ reaches values of the order of $Z$ for times 
$t \gg (Z/g)^{1/2}$. However, these fluctuations disappear in the limit $T \gg \epsilon_0$. 
For example, for $\zeta=10$ and ${\bar Z}=1$,  $z_G(t)-z_G(\infty)$ is of the order of $0.04 Z$ 
for $ t \simeq 40 (Z/g)^{1/2}$ at $T=5\epsilon_0$ but reaches again values of the order of 
$0.02 Z$ only for $ t \simeq 110(Z/g)^{1/2}$ at $T=10\epsilon_0$. The dependence on $\zeta$ 
and ${\bar Z}$ of the center of mass time evolution \eqref{zGfinal} is not simple, see 
Fig.\ref{fig:zGtphiZ}. We observe that the number of bounces increases with $Z$ and that 
the boson and fermion curves are very close to each other already for $\zeta=0.1$.
  
To obtain the large $N$ limit of the density \eqref{dens}, we write it in a form similar to 
\eqref{zG}-\eqref{J}.  In this case, the terms $p=q$ are present in the sum. Their contribution 
to the analog of the function $J$ is proportional to $\delta (\omega)$ and hence gives rise 
to a static component of the gas density wich we denote by $\rho_\infty$. The other terms 
can be evaluated with the help of the approximations used above to derive \eqref{zGfinal}. 
We find
\begin{eqnarray}   
\rho (z,t) &\simeq&  \rho_\infty (z) + \frac{4}{\Lambda} \sqrt{\pi{\bar Z}} \sum_{s > 0} s 
\int_0^{s\pi X} \frac{dx}{x^2}  \cos \left( x \sqrt{ \frac{g}{2 Z} } t \right)  \nonumber \\ 
&& \times  \frac{\sin  \sqrt{ (s\pi)^2 - x^2 }}{\sqrt{ (s\pi)^2 - z x^2/Z }} 
\frac{\cos  \sqrt{ (s\pi)^2 - z x^2/Z }}{e^{{\bar Z} [ ( s\pi/x )^2 - 1]} / \zeta \mp 1} \label{rhodyn}
\end{eqnarray}
where $X=\min (1,\sqrt{Z/z})$. For the asymptotic particle number density, we obtain
\begin{eqnarray}   
\rho_\infty (z) \simeq \frac{1}{\sqrt{\pi}\Lambda} 
\int_{\bar Z}^\infty \frac{dx}{\sqrt{x}} \frac{\Theta \big( x-mgz/T \big)}{\sqrt{x-mgz/T}} 
\frac{\big( x-{\bar Z} \big)^{1/2}}{e^{x-{\bar Z}}/\zeta \mp 1}  . 
\label{rhoinfty}
\end{eqnarray}
The large $N$ expressions \eqref{rhodyn} and \eqref{rhoinfty} agree very well with 
the exact results already for $T/\epsilon_0=50$, see Fig.\ref{fig:rhozt}. For $Z=0$, 
\eqref{rhoinfty} simplifies to \eqref{rhoz0} with $Z=0$. In this case, the density $\rho$ 
remains constant and equal to the thermal equilibrium density profile determined by 
the Hamiltonian $H_0$, energy $E$ and number $N$. For $Z \ne 0$, the asymptotic density 
profile is very different from the thermal equilibrium one. For $z \gg T/mg$, it coincides with 
the initial density \eqref{rhoz0}. For $z<Z$, the $\Theta$ function in \eqref{rhoinfty} can be 
replaced by $1$ and the derivative $\partial_z \rho_\infty$ is obviously positive and diverges 
for $z=Z$. The density $\rho_\infty$ thus assumes a maximum at a height $z>Z$, see 
Fig. \ref{fig:rhozt}. This conclusion also applies to the 3D case mentioned at the end of 
\ref{Pgcg}. For a Bose gas below the condensation temperature, the behavior of the condensate 
is clearly different from that of the thermal cloud, see Fig. \ref{fig:BEC} where the condensate 
fraction is $0.1$. The particle number density of the condensate at times multiple of 
$2(2Z/g)^{1/2}$ is essentially identical to its initial density \cite{Bongs} whereas 
the non-condensed gas relaxes to its asymptotic profile. 

\section{Influence of an environment \label{Ie}} 

Here, we compare the result of the relaxation process in the presence and absence 
of a heat bath. 
If the gas is coupled to a heat bath, the complete system consisting of the gas and 
its environment is described by a product Hilbert space 
${\cal H}_{gas} \otimes {\cal H}_{env}$ and by a Hamiltonian of the form $H=H_0+H_B+H_I$ 
where $H_B$ is the bath Hamiltonian and $H_I$ the interaction between the gas and 
the bath. In other words, the gas can exchange only energy with its environment. Owing to 
the coupling $H_I$, the state of the gas, initially given by \eqref{psi}, becomes mixed. 
To discuss the influence of the bath, we rewrite the initial state 
$\Omega=|\Psi \rangle \langle \Psi |$ as 
\begin{equation}
\Omega = \sum_{a,b} \Psi_a \Psi^*_b |a \rangle \langle b | \quad , \quad   
\Psi_a = \sum_{|\alpha\rangle \in {\cal H}_E} \Psi_\alpha \langle a |\alpha \rangle  
\end{equation} 
where we denote by Latin letters the $N$-particle eigenstates of the Hamiltonian $H_0$. 
Two cases must be distinguished. In the special case where $H_0$ and $H_I$ commute, 
the gas Hamiltonian is a constant of motion and hence the populations $|\Psi_a|^2$ remain 
constant. In this case, the effect of the bath is generically to destroy the coherences 
\cite{Endo1} and the gas state turns asymptotically into 
\begin{equation}
\Omega_\infty = \sum_{a} |\Psi_a|^2 |a \rangle \langle a | .
\end{equation} 
The corresponding particle number density $\rho_\infty (z)$ is equal to the constant component 
of \eqref{rho}. In the general case, $[H_I,H_0] \ne 0$, provided the coupling to the bath is 
weak enough, the asymptotic gas state is 
\begin{equation}
\Omega_\mathrm{can} = Z^{-1} \sum_{a} e^{-E_a/T_B} |a \rangle \langle a | 
\end{equation} 
where $Z=\sum_a \exp(-E_a/T_B)$, $E_a$ is the eigenenergy corresponding to $|a \rangle$ 
and $T_B$ is the bath temperature \cite{QDS,EPJB}. The gas relaxes into thermal equilibrium 
at the temperature of the bath. 

If the gas is isolated, its state remains pure but, as seen in the previous section, the particle 
number density can relax into the steady component $\rho_\infty$ of \eqref{rho}. Therefore, 
the asymptotic density profile of the isolated gas is not modified by an energy-conserving 
environment, i.e, such that $[H_I,H_0]= 0$. However, the relaxation towards this density is 
obviously different and depends on the details of the Hamiltonians $H_B$ and $H_I$. 
In the general case, if the coupling to the bath is weak enough, the density profile $\rho_\infty$ 
determined by the initial state $| \Psi \rangle$ can be observed in a transient regime before 
the bath imposes its temperature on the gas.

\section{Conclusion}

In this paper, we have studied an isolated perfect quantum gas confined by gravity and initially 
trapped above a certain height. The gas was assumed to be in a pure state of macroscopically 
well-defined energy and to evolve under Schr\"odinger dynamics. We have derived the expression 
for the time-dependent gas density profile in the limit of a large particle number. This single-particle distribution was found to depend not on the exact microscopic state of the gas but only on a few thermodynamic parameters:  the characteristic height of the initial confinement, the microcanonical
temperature and chemical potential corresponding to the gas energy and particle number. Damped
oscillations of the density profile were obtained. This relaxation behavior is a manifestation of 
the many-body nature of the system. The time evolution of a single confined particle is radically 
different. As an important consequence of this difference, the time-dependent density of a Bose 
gas below the condensation temperature consists of two clearly distinguishable components 
corresponding to the condensate and the thermal cloud. However, though the gas density relaxes, 
in general, towards a well-determined profile, this asymptotic density is very different from 
the thermal equilibrium one determined by the gas energy and particle number. Contrary to 
the latter, the former depends also on the initial height of the gas and presents a maximum.  

We have also derived general results about the time-evolution of an isolated many-body system. 
We have shown that, the single-particle distributions of a typical pure state of macroscopically 
well-defined energy are identical to that of the microcanonical mixed state at the same energy. 
We stress that this equivalence holds for any many-body system, our proof does not depend on 
the absence or presence of particle-particle interactions. Moreover, as the single-particle 
distributions are the same for almost all pure states of a given macroscopic energy, 
the microcanonical results should be valid for a large class of statistical ensembles and apply 
to experimental measures obtained with different systems prepared ''under identical 
experimental conditions''. A coupling of the system to environmental degrees of freedom 
modifies the time evolution of the single-particle distributions. However, as discussed in 
the previous section, the asymptotic distributions found assuming the system is truly isolated 
are not changed by an energy-conserving environment and could be observed in a transient 
regime if energy is exchanged with the environment. 

We finally comment about the particle-particle interactions. We first emphasize that our results 
show that these interactions are not necessary to obtain a relaxation behavior of physically 
relevant degrees of freedom of a many-body system. As we have seen, this behavior results 
from the properties of the microcanonical single-particle density matrix. Furthemore, the dilute 
regime is experimentally accessible. In this regime, for Bose gases, the interactions cannot be 
neglected below the condensation temperature as the condensate size is essentially fixed by 
the confining potential but they might play a minor role above this temperature \cite{revue}. 
The theoretical study of the influence of interactions between the gas particles and other issues 
are left for future work.

\begin{acknowledgments}
We thank R. Chitra, K. Kruse and L. Pricoupenko for stimulating discussions on related topics.
\end{acknowledgments}

\section*{Appendix} 
 
To evaluate the scalar product \eqref{spqk}, we write
\begin{eqnarray}   
&&\int_{X}^\infty dx A(x+a_q) \left[  \partial_x^2 -x \right] A \left( x-X+a_k \right) \\
&&\phantom{\int_{X}^\infty d}= \left( a_k-X \right) \int_X^\infty dx A(x+a_q) 
A \left( x-X+a_k \right) \nonumber  \\
&&\phantom{\int_{X}^\infty d}= -A\left( X+a_q \right) \partial_x A(a_k) \\ 
&&\phantom{\int_{X}^\infty dx A}+ a_q    \int_X^\infty dx A(x+a_q) 
A \left( x-X+a_k \right)  \nonumber 
\end{eqnarray} 
where $X=Z/z_0$. 
The first equality is obtained with the help of the equation $[\partial_x^2 - x]A(x)=0$ 
and the second one using this equation and integrations by parts. As $A(a_q)=0$, 
the above result for $X=0$ gives the orthogonality of the wavefunctions $\phi_k$ and 
$\phi_q$ and, for $k=q$,  the equality of the derivatives with respect to $X$ at $X=0$ 
of both sides gives the normalization constant of $\phi_k$.      

By integrations by parts and using the fact that $\phi_q$ are the one-particle 
eigenfunctions of the Hamiltonian \eqref{H} with $Z=0$, the time derivative of the gas 
center of mass \eqref{zGdef} can be written as
\begin{equation}   
\partial_t z_G  = \frac{i}{Nm} \sum_{p,q} \langle c^{\dag}_{p} c^{\phantom{\dag}}_{q} \rangle 
e^{it (\epsilon_p - \epsilon_q )} \int_0^\infty dz \phi_q (z)  \partial_z \phi_p  .
\end{equation}
This expression is valid for any confining potential and simply states that the velocity 
of the gas center of mass is the average momentum of the gas divided by its mass 
$Nm$. In the same manner, we find
\begin{equation}   
\partial_t^2 z_G  =  \frac{g}{N} \sum_{p \ne q} \langle c^{\dag}_{p} c^{\phantom{\dag}}_{q} \rangle 
e^{it (\epsilon_p - \epsilon_q )}
\end{equation}   
where we have used the normalization of the density \eqref{dens} and 
$\partial_z \phi_p (0^+) = (2m^2 g)^{1/2}$ for any $p$. 
As $\langle c^{\dag}_{p} c^{\phantom{\dag}}_{q} \rangle
=\langle c^{\dag}_{q} c^{\phantom{\dag}}_{p} \rangle$ 
is real, $\sum_{p \ne q} \langle c^{\dag}_{p} c^{\phantom{\dag}}_{q} \rangle
/(\epsilon_p - \epsilon_q )=0$ and $\partial_t z_G(0)=0$. Consequently, the center of mass 
$z_G$ is given by \eqref{zG}.

At initial time, the gas density vanishes at $z=0$ and hence we expect an acceleration 
$\partial_t^2 z_G (0) =  -g$. This can be shown in the following way. Let us consider 
a wavefunction $\varphi(z)$ which vanishes for $z<h$ where $h > 0$. We expand it on 
the basis $\{ \phi_p \}$ and write its derivative at $z=0$ as
\begin{equation}   
\partial_z \varphi (0) = (2m^2 g)^{1/2} \int_0^\infty dz \varphi(z) \sum_p \phi_p (z) = 0 .
 \end{equation}   
We deduce from this property of the sum $\sum_p \phi_p (z)$ that 
$\sum_{p,q} \langle c^{\dag}_{p} c^{\phantom{\dag}}_{q} \rangle=0$ where 
$\langle c^{\dag}_{p} c^{\phantom{\dag}}_{q} \rangle$ is given by \eqref{cpcq} and 
\eqref{spqk}. Therefore, $\sum_{p \ne q} \langle c^{\dag}_{p} c^{\phantom{\dag}}_{q} \rangle
=- \sum_{p} \langle c^{\dag}_{p} c^{\phantom{\dag}}_{p} \rangle=-N$ and we obtain 
the expected initial acceleration.

\end{document}